\documentclass{article}
\usepackage{spconf,amsmath,graphicx,hyperref}
\usepackage{array}
\usepackage{multirow}
\usepackage{amssymb}
\newcolumntype{C}[1]{>{\centering\arraybackslash}p{#1}}

\usepackage{etoolbox}
\makeatletter
\pretocmd{\thebibliography}{
    \begingroup
    \linespread{0.98}\selectfont
    \small
}{}{}
\apptocmd{\thebibliography}{%
  \setlength{\itemsep}{-0.1ex}%
  \setlength{\parskip}{0pt}%
}{}{}
\apptocmd{\endthebibliography}{\endgroup}{}{}
\makeatother

\title{MEANFLOWSE: ONE-STEP GENERATIVE SPEECH ENHANCEMENT VIA CONDITIONAL MEAN FLOW}

\name{Duojia Li$^{\star}$, Shenghui Lu$^{\dagger}$, Hongchen Pan$^{\star}$,
      Zongyi Zhan$^{\star}$, Qingyang Hong$^{\dagger\ast}$, Lin Li$^{\star\ast}$%
\thanks{*Corresponding authors: Lin Li and Qingyang Hong.\protect\\
\hspace*{1.5em}This work was supported in part by the National Natural Science
Foundation of China under Grants 62371407 and 62276220.}
}

\address{$^{\star}$ School of Electronic Science and Engineering, Xiamen University, China\\
$^{\dagger}$ School of Informatics, Xiamen University, China\\
liduojia@stu.xmu.edu.cn; qyhong@xmu.edu.cn; lilin@xmu.edu.cn}

\begin{document}
%
\maketitle
\begin{abstract}
Multistep inference is a bottleneck for real-time generative speech enhancement because flow and diffusion-based systems learn an instantaneous velocity field and therefore rely on iterative ordinary differential equation solvers. We introduce MeanFlowSE, a conditional generative model that learns the average velocity over finite intervals along a trajectory. Using a Jacobian–vector product to instantiate the MeanFlow identity, we derive a local training objective that directly supervises finite-interval displacement while remaining consistent with the instantaneous-field constraint on the diagonal. At inference, MeanFlowSE performs single-step generation via a backward-in-time displacement, removing the need for multistep solvers; an optional few-step variant offers additional refinement. On VoiceBank–DEMAND, the single-step model achieves strong intelligibility, fidelity, and perceptual quality with substantially lower computational cost than multistep baselines. The method requires no knowledge distillation or external teachers, providing an efficient, high-fidelity framework for real-time generative speech enhancement. The proposed method is open-sourced at \url{https://github.com/liduojia1/MeanFlowSE}.

\end{abstract}
\section{Introduction}
\label{sec:intro}

Speech enhancement (SE) aims to recover clean speech from noisy signals, playing a vital role in communication systems and robust automatic speech recognition (ASR)~\cite{loizou_book,mmse_stsa}. Discriminative methods estimate spectral masks or clean-speech features~\cite{dccrn,convtasnet}, yet in adverse conditions they can produce over-smoothed or distorted outputs, reducing perceptual quality and intelligibility~\cite{segan}. Recent deterministic architectures such as ZipEnhancer~\cite{wang2025zipenhancer} and Mamba-SEUNet~\cite{wang2025mamba} have also shown strong performance for monaural SE.

\begin{figure}[!t]
  \centering
  \includegraphics[width=\columnwidth]{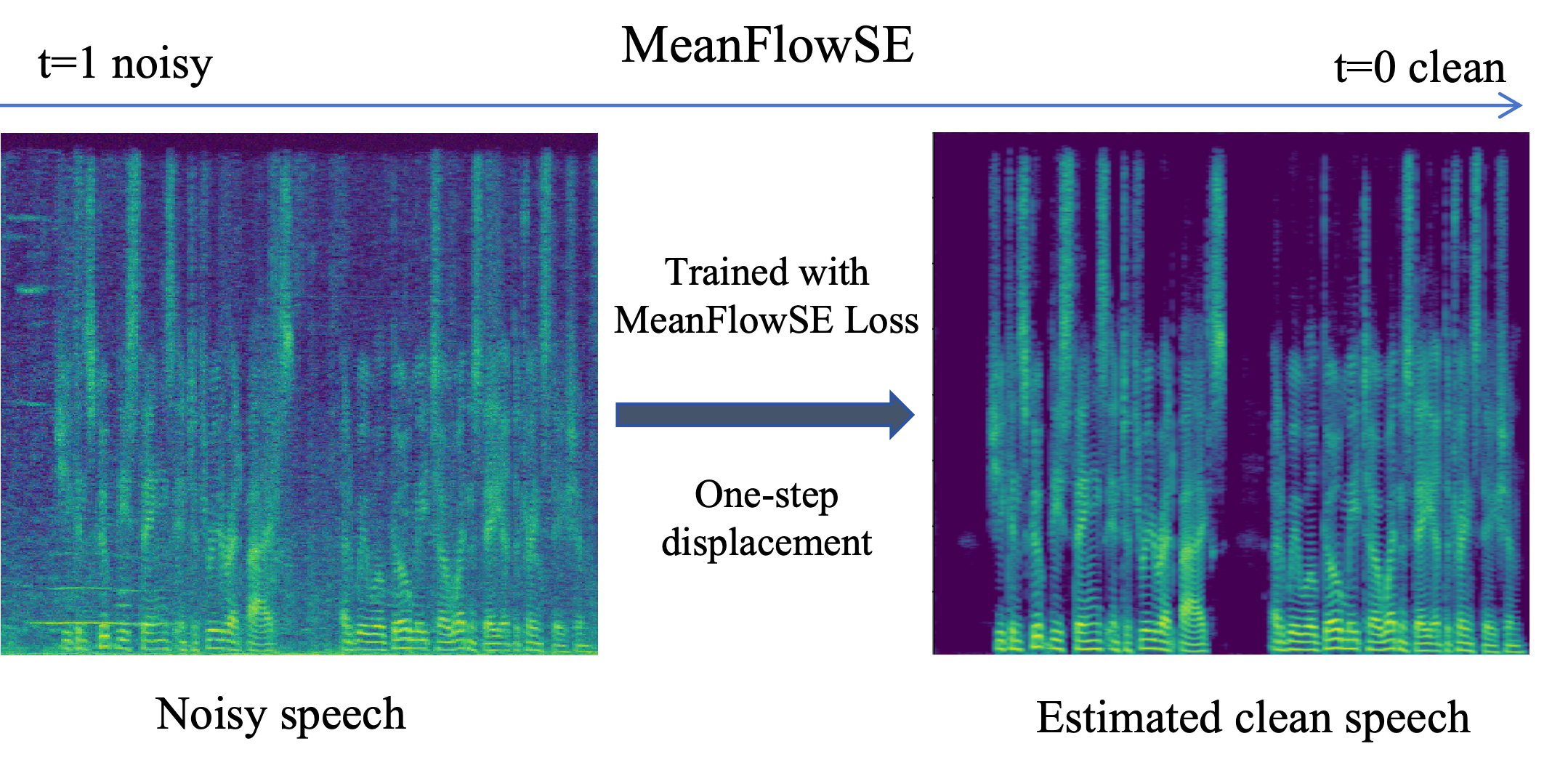}
  \caption{One-step backward-in-time displacement. Trained with the MeanFlowSE loss, the model maps the noisy spectrogram at $t{=}1$ to an enhanced estimate via a single finite-interval displacement along the conditional path toward $t{=}0$}
  \label{fig:euler_mf}
\end{figure}

Generative models provide an alternative by learning the clean–speech distribution and inverting the noise corruption process~\cite{ddpm,sde_score}. Diffusion or score-based methods in the short-time Fourier transform (STFT) domain have achieved strong performance, yet their reliance on numerous function evaluations (NFE) limits real-time applicability. Several efforts aim to mitigate this issue: CDiffuSE employs conditional reverse sampling to improve fidelity, yet still relies on long sampling chains~\cite{cdiffuse}; SGMSE stabilizes complex spectral synthesis with predictor–corrector methods, but retains high computational cost~\cite{sgmse_complex}; Correcting the reverse process (CRP) learns a correction term to reduce sampling steps, albeit requiring additional fine-tuning~\cite{rpc}; the Schr\"odinger Bridge (SB) formulation treats SE as a stochastic control problem, yet remains multi-step and sensitive to discretization~\cite{sbridge}. Among normalizing flow techniques, flow matching (FM) offers a deterministic alternative counterpart to diffusion~\cite{normalizing_flows,flowmatching}. Conditional flow matching (CFM) further regulates the instantaneous velocity field along a prescribed path~\cite{flowmatching}. FlowSE implements CFM for speech with a linear–Gaussian conditional path achieving performance comparable to diffusion models with fewer steps; yet it still depends on iterative numerical integration of the instantaneous velocity, falling short of efficient one-step inference~\cite{flowse}.

In this paper, we propose MeanFlowSE, a generative speech enhancement model that learns an average‑velocity field, capturing finite‑interval displacement rather than an instantaneous slope. Our approach employs the MeanFlow identity~\cite{meanflows} using a Jacobian–vector product objective, specifically adapted for conditional speech enhancement. This objective supervises the finite‑interval average field while naturally reducing to standard CFM on the diagonal ($r{=}t$). During inference, the learned displacement replaces iterative numerical ordinary differential equation (ODE) integration, enabling one‑step generation and few‑step refinement under a unified framework (Fig.~\ref{fig:euler_mf}). Evaluated on VoiceBank-DEMAND, MeanFlowSE achieves performance competitive with or superior to strong flow and diffusion baselines, while operating at a significantly lower real-time factor (RTF). The model is trained from scratch without knowledge distillation~\cite{progressive_distillation} and remains compatible with inference-acceleration techniques from related domains, such as rectified flows and consistency models~\cite{rectified_flow,consistency_models}.

\section{RELATED WORK}
\label{sec:background}

\subsection{FlowSE}
FlowSE~\cite{flowse} learns a deterministic instantaneous velocity field by conditional flow matching on a prescribed linear–Gaussian path. Let $x_1$ denote the clean speech signal and $y$ the corresponding noisy observation. A time variable $t\in[0,1]$ parameterizes the interpolation from $y$ to $x_1$ through the linear schedules:
\begin{align}
  \mu_t(x_1,y)&=t\,x_1+(1-t)\,y, \label{eq:mu_t}\\
  \sigma_t&=(1-t)\,\sigma, \label{eq:sigma_t}
\end{align}
where $\sigma>0$ controls the noise level. Here, $\mu_t$ defines the path mean between $y$ and $x_1$, while $\sigma_t$ specifies the standard deviation of the Gaussian perturbation used to sample $x_t$ around $\mu_t$. Along this path, differentiating the state with respect to $t$ and eliminating the auxiliary variable yields the closed-form on-path instantaneous velocity target:
\begin{equation}
  v_t(x_t\mid x_1,y)=\frac{\sigma_t'}{\sigma_t}\big(x_t-\mu_t\big)+\mu_t'
  =\frac{x_1-x_t}{1-t},
  \label{eq:teacher}
\end{equation}
with $\mu_t'=x_1-y$ and $\sigma_t'=-\sigma$. The conditional flow-matching loss regresses a network $v_\theta(x_t,t,y)$ to the target:
\begin{equation}
  \mathcal{L}_{\mathrm{CFM}}
  =\mathbb{E}_{(x_1,y),\,t}\!\left[\big\|\,v_\theta(x_t,t,y)-v_t(x_t\mid x_1,y)\,\big\|_2^2\right],
  \label{eq:cfm_loss}
\end{equation}
where $x_t$ lies on the path defined by Eqs.\eqref{eq:mu_t} and \eqref{eq:sigma_t} and $t$ is sampled uniformly on $[0,1-\delta]$ for a small $\delta>0$ to avoid the singular limit $t\!\to\!1$. Inference initializes from the noisy-side prior $x_0\sim\mathcal{N}(y,\sigma^2 I)$ and proceeds on a grid $0=t_0<\cdots<t_N=1$ by the explicit Euler update:
\begin{equation}
  x_{t_i}=x_{t_{i-1}}+v_\theta\!\big(x_{t_{i-1}},t_{i-1},y\big)\,\Delta t(i),
  \label{eq:flowse_sample}
\end{equation}
where the step size is defined as $\Delta t(i)=t_i-t_{i-1}$. This rule implements the forward integration of the learned vector field along the interpolation path.

\subsection{Mean Flows}
\label{ssec:meanflows}
Mean Flows~\cite{meanflows} introduces the average velocity field to describe finite-interval motion, 
in contrast to the instantaneous velocity field used in conventional flow matching. 
Let the dynamics follow the marginal instantaneous field $v(z_t,t)$, i.e.,  
$z_t' = v(z_t,t)$, where $z_t\!\in\!\mathbb{C}^{D}$ denotes the interpolation path at time $t\!\in\![0,1]$, and $(\cdot)'$ is the derivative with respect to $t$. For any $r<t$, the average velocity is defined by:
\begin{equation}
  u(z_t,r,t) = \frac{1}{t-r}\int_r^t v(z_\tau,\tau)\,d\tau,
  \label{eq:avg_vel}
\end{equation}
where displacement over elapsed time yields the definition 
$u(z_t,t,t)=v(z_t,t)$. Differentiating $(t-r)u$ with respect to $t$ gives the 
MeanFlow identity:
\begin{equation}
  u(z_t,r,t) = v(z_t,t) - (t-r)\,\frac{d}{dt}u(z_t,r,t),
  \label{eq:meanflow_id}
\end{equation}
where the total derivative is expanded by the chain rule~\cite{meanflows} as:
\begin{equation}
  \frac{d}{dt}u(z_t,r,t) 
  = v(z_t,t)\,\partial_z u + \partial_t u.
  \label{eq:chainrule}
\end{equation}

This identity yields a computable local supervision rule. Parameterizing the average 
field with a network $u_\theta$, and holding $r$ fixed at $(z_t,t)$, one forms the 
regression target:
\begin{equation}
  u_{\mathrm{tgt}} = v(z_t,t) - (t-r)\Big(v(z_t,t)\,\partial_z u_\theta
  + \partial_t u_\theta\Big),
  \label{eq:utgt}
\end{equation}
and the loss with stop-gradient $\mathrm{sg(\cdot)}$ on the target:
\begin{equation}
  \mathcal{L} = \mathbb{E}\!\left[\left\|\,u_\theta(z_t,r,t) - \mathrm{sg}(u_{\mathrm{tgt}})\,\right\|_2^2\right].
  \label{eq:meanflow_loss}
\end{equation}

Once $u_\theta$ is learned, 
sampling replaces ODE integration with a displacement rule:
\begin{equation}
  z_r = z_t - (t-r)\,u_\theta(z_t,r,t),
  \label{eq:displacement}
\end{equation}

Mean Flows provides both a training objective and a displacement-based sampling rule in Eq.~\eqref{eq:displacement}~\cite{meanflows}. This formulation transports the state from $t$ to an earlier $r$ in a single update and recovers Euler integration in the small-step limit.

\section{Method}
\label{sec:method}

Motivated by the Mean Flows principle, we propose to learn an average velocity field for conditional speech enhancement.
The proposed method operates in the complex STFT domain with paired samples $(x_1,y)$, where $x_1$ denotes the clean speech signal and $y$ the corresponding noisy observation. To ensure consistency between training and inference, we adopt the dual linear--Gaussian 
conditional path:
\begin{align}
  \mu_t &= (1-t)\,x_1 + t\,y, \label{eq:mu_method} \\
  \sigma_t &= (1-t)\,\sigma_{\min} + t\,\sigma_{\max}, \label{eq:sigma_method}
\end{align}
where $t\in[0,1]$, so that $t=0$ is the clean endpoint
$(\mu_0=x_1,\ \sigma_0=\sigma_{\min})$ and $t=1$ is the noisy endpoint
$(\mu_1=y,\ \sigma_1=\sigma_{\max})$. This dual parameterization reverses the endpoint convention of FlowSE. 
Training points are drawn on the path via:
$\  x_t \;=\; \mu_t + \sigma_t\, z$, where $z\sim\mathcal{N}(0,I)$.

Differentiation along
the path gives the on-path instantaneous target:
\begin{equation}
  v_t(x_t \mid x_1,y) 
  = \mu_t' + \sigma_t' z 
  = \frac{\sigma_t'}{\sigma_t}\,(x_t-\mu_t)+\mu_t',
  \label{eq:teacher_method}
\end{equation}
with $\mu_t' = y-x_1$ and $\sigma_t'=\sigma_{\max}-\sigma_{\min}$. This target is used
only at sampled points on the path.

\subsection{Average velocity and the MeanFlowSE identity}
\label{ssec:meanflowse-identity}
Iteratively integrating a local slope can accumulate error on curved trajectories.
Instead, we estimate the finite-interval average velocity, defined as the constant rate that yields the net displacement between two time points.

Let $v(x,t\mid y)$ denote the marginal instantaneous field that governs the
conditional dynamics $x_t' = v(x_t,t\mid y)$.
For any $r<t$, define the average velocity as the mean slope that reproduces the
displacement on $[r,t]$:
\begin{equation}
  u(x_t,r,t\mid y)=\frac{1}{t-r}\int_{r}^{t} v(x_\tau,\tau\mid y)\,d\tau,
  \label{eq:avgvel_method}
\end{equation}
so $u(x_t,t,t\mid y)=v(x_t,t\mid y)$.
Differentiating $(t-r)u$ with respect to $t$ yields the MeanFlowSE identity:
\begin{equation}
  u(x_t,r,t\mid y)=v(x_t,t\mid y)-(t-r)\frac{d}{dt}u(x_t,r,t\mid y),
  \label{eq:mfse_identity}
\end{equation}
where the total derivative is taken along the conditional trajectory:
\begin{equation}
  \frac{d}{dt}u(x_t,r,t\mid y)
  = v(x_t,t\mid y)\!\cdot\!\nabla_x u
  + \partial_t u.
  \label{eq:total_derivative_method}
\end{equation}

The intractable path integral in Eq.\eqref{eq:avgvel_method} is replaced in Eqs.~\eqref{eq:mfse_identity}-\eqref{eq:total_derivative_method} by local terms evaluated at $(x_t,t)$, where the diagonal case $r=t$ simplifies to $u=v$.

\subsection{MeanFlowSE loss}
\label{ssec:mfse-loss}
The core idea of our approach is to train a single network
$u_\theta(x,r,t,y)$ so that it satisfies
the identity given in Eqs.~\eqref{eq:mfse_identity}-\eqref{eq:total_derivative_method}
at training samples $(x_t,r,t)$ drawn from
the path in Eqs.~\eqref{eq:mu_method}-\eqref{eq:sigma_method}.
By substituting the closed-form target $v_t$ from Eq.~\eqref{eq:teacher_method} into Eq.~\eqref{eq:mfse_identity}, and then expanding the total derivative in Eq.~\eqref{eq:total_derivative_method} with $(y,r)$ held fixed, we obtain the first-order training objective:
\begin{equation}
  u_{\mathrm{tgt}}=v_t
  - c\,(t-r)\!\left[v_t\!\cdot\!\nabla_x u_\theta
  + \partial_t u_\theta\right],
  \label{eq:utgt_method}
\end{equation}

The first order target in Eq.~\eqref{eq:utgt_method} coincides with the Mean Flow identity when $c=1$; we adopt $c=0.5$ as a stabilizing first order correction to enhance stability without altering the fixed point. Motivated by the original Mean Flow framework, a stop-gradient operation is applied to the target to prevent higher-order backpropagation through the Jacobian–vector product term.

We train the network to approximate this target by minimizing the MeanFlowSE loss with stop-gradient, which avoids higher-order backpropagation and yields a well-posed fixed point:
\begin{equation}
  \mathcal{L}_{\mathrm{MFSE}}
  = \mathbb{E}\!\left[\left\|
    u_\theta(x_t,r,t,y) - \mathrm{sg}(u_{\mathrm{tgt}})
  \right\|_2^2\right].
  \label{eq:lmfse}
\end{equation}

At the boundary $r=t$, the target in Eq.\eqref{eq:utgt_method} reduces to
$u_{\mathrm{tgt}}=v_t$, causing Eq.\eqref{eq:lmfse} to coincide exactly with the
conditional flow matching objective along the diagonal. In practice, we incorporate a small
fraction of boundary samples during training. Derivatives are computed using automatic differentiation, supplemented by a numerically stable
centered finite-difference method as a fallback; only $(x,t)$ are differentiated, while
$(y,r)$ are treated as constants.

\subsection{One-step inference}
\label{ssec:onestep}
Since the network learns a finite-interval displacement field, inference no longer requires integrating instantaneous velocities. We apply a backward-in-time Euler step update that directly transports the noisy endpoint to the enhanced estimate. After training, numerical ODE integration is replaced by a displacement driven by the learned field:
\begin{equation}
  x_{t_{k+1}} \;=\; x_{t_k} \;-\; \Delta_k\,
  u_\theta\!\left(x_{t_k},\, r=t_{k+1},\, t=t_k \mid y\right),
  \label{eq:displace_step}
\end{equation}
where $k$ denotes the time-step index ($k=0,\dots,N-1$), $\Delta_k=t_k-t_{k+1}>0$, and $\{t_k\}_{k=0}^N$ is a monotone decreasing grid $T_{\mathrm{rev}}=t_0>t_1>\cdots>t_N=t_\varepsilon$. 

The reverse-time initialization uses the noisy-side marginal $x_{T_{\mathrm{rev}}}\sim\mathcal{N}\!\big(y,\sigma^2(T_{\mathrm{rev}})I\big)$,
where $T_{\mathrm{rev}}\in(0,1]$ is the reverse-time start and $t_\varepsilon\in[0,1)$ is the terminal time used for numerical stability.

A single-step inference rule follows:
\begin{equation}
  \hat{x}_{t_\varepsilon}
  \;=\; x_{T_{\mathrm{rev}}}
  \;-\; (T_{\mathrm{rev}}-t_\varepsilon)\,
    u_\theta\!\left(x_{T_{\mathrm{rev}}},\, r=t_\varepsilon,\, t=T_{\mathrm{rev}}\mid y\right).
  \label{eq:single_step}
\end{equation}

\begin{table*}[!t]
  \centering
  {\scriptsize\caption{\centering \MakeUppercase{Performance comparisons with the state-of-the-art speech enhancement systems on the VoiceBank–DEMAND dataset.}}\label{tab:sota}}
  \footnotesize
  \setlength{\tabcolsep}{5pt}
  \renewcommand{\arraystretch}{1.05}
  \begin{tabular*}{\textwidth}{@{\extracolsep{\fill}} C{32mm} *{9}{c}}
    \hline
    \multirow{2}{*}{\parbox[c]{32mm}{\centering\textbf{System}}} &
    \multirow{2}{*}{\textbf{NFE}} &
    \multicolumn{3}{c}{\textbf{DNSMOS}} &
    \multirow{2}{*}{\textbf{PESQ} $\uparrow$} &
    \multirow{2}{*}{\textbf{ESTOI} $\uparrow$} &
    \multirow{2}{*}{\textbf{SI-SDR} $\uparrow$} &
    \multirow{2}{*}{\textbf{SpkSim} $\uparrow$} &
    \multirow{2}{*}{\textbf{RTF} $\downarrow$} \\
    \cline{3-5}
    & & \textbf{SIG} $\uparrow$ & \textbf{BAK} $\uparrow$ & \textbf{OVRL} $\uparrow$ & & & & & \\
    \hline
    Noisy                   & - & 3.346 & 3.126 & 2.697 & 1.970 & 0.787 &  8.445 & 0.888 & - \\
    SGMSE                   & 30  & \textbf{3.488} & 3.985 & 3.176 & 2.922 & 0.863 & 17.396 & \underline{0.891} & 1.81 \\
    FlowSE                  & \underline{5}   & 3.478 & 4.051 & 3.202 & \textbf{3.047} & \underline{0.873} & 19.145 & 0.889 & \underline{0.23} \\
    Schr\"odinger Bridge  & 30  & 3.486 & \underline{4.062} & \textbf{3.216} & 2.901 & 0.872 & \underline{19.448} & 0.886 & 1.07 \\
    CDiffuSE                & 200 & 3.434 & 3.727 & 2.994 & 2.513 & 0.798 & 13.665 & 0.812 & 6.94 \\
    StoRM                   & 50  & \underline{3.487} & 4.031 & 3.204 & 2.891 & 0.868 & 18.518 & \underline{0.891} & 2.61 \\
    MeanFlowSE (Ours)       & \textbf{1} & 3.471 & \textbf{4.073} & \underline{3.207} & \underline{2.942} & \textbf{0.881} & \textbf{19.975} & \textbf{0.892} & \textbf{0.11} \\
    \hline
  \end{tabular*}
    {\par\vspace{0.25em}\raggedright\scriptsize\textit{Note:} In this paper, bold indicates the best score in each column and underline indicates the second best.\par}
\end{table*}
\begin{table}[t]
  \centering
  \caption{\centering \MakeUppercase{Quality–Efficiency Trade-off: FlowSE vs. MeanFlowSE}}
  \label{tab:nfe_single}
  \scriptsize
  \setlength{\tabcolsep}{5pt}
  \renewcommand{\arraystretch}{1.05}
  \begin{tabular*}{\columnwidth}{@{\extracolsep{\fill}} c c c c c c}
    \hline
    \textbf{System} & \textbf{NFE} & \textbf{ESTOI} $\uparrow$ & \textbf{SI-SDR} $\uparrow$ & \textbf{SpkSim} $\uparrow$ & \textbf{RTF} $\downarrow$ \\
    \hline
    FlowSE            & 1  & 0.872            & \underline{19.560}  & 0.880            & \underline{0.16} \\
    FlowSE            & 5  & \underline{0.873} & 19.145            & 0.889            & 0.23 \\
    FlowSE            & 10 & 0.870            & 18.428            & \underline{0.891} & 0.38 \\
    FlowSE            & 20 & 0.868            & 18.099            & 0.890            & 0.71 \\
    MeanFlowSE (Ours) & \textbf{1} & \textbf{0.881} & \textbf{19.975} & \textbf{0.892} & \textbf{0.11} \\
    \hline
  \end{tabular*}
\end{table}

\section{Experiments}
\label{sec:experiments}

We evaluate on the VoiceBank-DEMAND (VB-DMD) corpus at 16\,kHz using the standard splits, where validation speakers are held out and both test speakers and SNRs remain unseen~\cite{vb_demand1,vb_demand2}. All systems share the same STFT front end (Hann window, centered frames) and waveform normalization: signals are peak-normalized by the noisy input, and complex spectra are represented as $|z|^{0.5}\exp(j\angle z)$ with a global scale factor of 0.15.

Our enhancement network is based on NCSN++ with self-attention~\cite{ncsnpp,unet,attention}. We concatenate $(x_t,y)$ along the channel dimension and condition the model on Gaussian Fourier features of $t$ and $\Delta=t-r$, enabling it to predict a complex-valued vector field. Training minimizes the MeanFlowSE loss in Eq.~\eqref{eq:lmfse} by mixing diagonal samples ($\Delta=0$) and off-diagonal samples ($\Delta>0$) drawn from the dual path in Eqs.~\eqref{eq:mu_method}--\eqref{eq:sigma_method}. We train with Adam~\cite{adam} using a learning rate of $10^{-4}$, an exponential moving average (EMA) decay of 0.999, and gradient clipping at 1.0. To stabilize training, we adopt curriculum learning: we first train on instantaneous velocity, then gradually increase the weight of the average-velocity objective.

Baselines include CDiffuSE~\cite{cdiffuse}, SGMSE~\cite{sgmse_complex}, and StoRM~\cite{storm}. We also evaluate reverse-process correction~\cite{rpc}, Schr\"odinger Bridge~\cite{sbridge}, and FlowSE~\cite{flowse} following the authors' recommended settings. NFE denotes the number of forward passes per utterance. Metrics include PESQ~\cite{pesq}, ESTOI~\cite{estoi}, SI-SDR~\cite{sisdr}, DNSMOS P.835~\cite{dnsmos} (SIG/BAK/OVRL), and speaker similarity (SpkSim)~\cite{spksim}. We also report the real-time factor (RTF), measured end-to-end with batch size one on a single V100.

\section{Results}
\label{sec:results}

Table~\ref{tab:sota} presents a system-level comparison on the VB-DMD under an identical front end and normalization. With only one function evaluation, MeanFlowSE attains the highest overall quality, an ESTOI of 0.881 and SI-SDR of 19.975\,dB. It also attains the best BAK of 4.073, along with a competitive OVRL score of 3.207, a speaker similarity of 0.892, and a PESQ of 2.942. Notably, MeanFlowSE achieves the lowest RTF of 0.11 in the comparison. In contrast, prior systems based on diffusion, bridge, and flow require between 5 and 200 evaluation steps, yielding RTFs ranging from 0.23 to 6.94 under the same setup. Table~\ref{tab:nfe_single} further analyzes the quality--efficiency trade-off between FlowSE and MeanFlowSE by varying the number of function evaluations for FlowSE. MeanFlowSE achieves the best overall performance across the reported metrics while retaining the lowest real-time factor, even with a single function evaluation.
Direct supervision of finite-interval displacement reduces the error accumulation of multi-step integration over noisy instantaneous fields.

These results underscore the impact of the underlying modeling strategy. Methods such as SGMSE, StoRM, Schr\"odinger’s Bridge, CDiffuSE, and FlowSE integrate an instantaneous field with a multi-step ODE solver, where quality improvements come at the cost of increased inference time. In contrast, MeanFlowSE learns an average velocity field that directly predicts the finite-interval displacement and enables a single backward-in-time update during inference. By compressing the generative trajectory to just one step, MeanFlowSE maintains high signal fidelity and intelligibility while delivering the superior background noise suppression and the lowest computational cost. Overall, the results indicate that average-velocity learning advances the quality–efficiency frontier for generative speech enhancement without distillation or external teachers.

\section{Conclusion}
\label{sec:conclusion}
We propose MeanFlowSE, a single-step generative SE model that learns an average velocity field. Using the MeanFlow identity on a conditional path, we derive a tractable objective whose optimum matches the finite-interval displacement, enabling displacement-based inference without ODEs. MeanFlowSE achieves competitive quality against strong multistep baselines at much lower cost. Limitations include the linear--Gaussian path and first-order derivatives; future work will explore more flexible paths and real-world evaluation.

\makeatletter
\let\oldthebibliography\thebibliography
\let\endoldthebibliography\endthebibliography
\renewenvironment{thebibliography}[1]{%
  \oldthebibliography{#1}%
  \setlength{\itemsep}{-0.40ex}%
  \setlength{\parsep}{0pt}%
  \setlength{\parskip}{0pt}%
  \setlength{\topsep}{0pt}%
  \setlength{\partopsep}{0pt}%
  \linespread{0.92}\selectfont%
}{%
  \endoldthebibliography%
}
\makeatother


\bibliographystyle{IEEEbib}
\bibliography{strings,refs}

\end{document}